\documentstyle[12pt]{article}
\input epsf

\begin{document}

\begin{flushright} ULB--TH--98/03 \\ hep-th/9802142  \\
February 1998\\ \end{flushright}

\vspace{2cm}

\begin{center} {\LARGE Spontaneous Symmetry
Breaking in Gauge Theories: a Historical Survey} \footnote{
Talks presented at the Award  Ceremony of the 1997 High Energy
and Particle Physics Prize of the European Physical Society
(Jerusalem, 24 August 1997)}   \vspace{2cm}

{\large R.~Brout}\footnote{
  E-mail: rbrout@ulb.ac.be} $^{*}$ {\large and
F.~Englert}\footnote { E-mail: fenglert@ulb.ac.be}
$^{*,\dagger}$ \addtocounter{footnote}{-3}\\ \vspace{.4cm}
$^{*}${\it Service de Physique Th\'eorique}\\ {\it
Universit\'e Libre de Bruxelles, Campus Plaine, C.P.225}\\
{\it Boulevard du Triomphe, B-1050 Bruxelles, Belgium} \\
\vspace{.4cm}

$^{\dagger}${\it Institute of Advanced Studies}\\ {\it Hebrew
University}\\{\it Givat Ram, Jerusalem 91904, Israel}\\

\end{center}

\vspace{1.5cm}

\begin{abstract}

\noindent The personal and scientific history of the
discovery of spontaneous symmetry breaking in gauge theories is
outlined and its scientific content is reviewed.

\end{abstract}

\newpage

\noindent  {\bf   HISTORY }\footnote{Presented by F. Englert}
\addtocounter{footnote}{-1}

Our discovery of spontaneous symmetry breaking in gauge theory
is intimately linked to the history of our collaboration.
Evoking this period of our life, I shall survey the scientific
history of spontaneous symmetry breaking.  A more detailed
historical review  can be found  in the talk presented by
Veltman  at the International Symposium on Electron and Photon
Interactions at High Energies in 1973 at Bonn.

I came to Cornell University in 1959 as a Research Associate
to Robert Brout who was Professor there. My background was
mainly in solid state physics and many body problems, and
Robert Brout was already well known for his work in these
fields. Our first contact was unexpectedly   warm. It started
right when he came to take me at the airport with his near
century old Buick and took me to a drink which lasted up to
the middle of the night. When we left, we knew that we would
become friends.

At Cornell, we   realized that, in our approach to physics, we
were   different.  Robert had an amazing easiness in
translating   abstract concepts into tangible intuitive
images, whereas I, with my latin oriented education, had on
the contrary always a tendency to express images in terms of
formal structures. But this difference turned into a fruitful
complementarity because   we quickly learned to understand the
functioning of each others mind. Still today, talking physics,
each of us gets somehow frustrated not to be able to terminate
a sentence, as the other does it for him, and is apparently
very happy to do so.

Playing together on many-body problems, we got involved  in
the study of phase transitions and particularly in
ferromagnetism. We understood the importance of the spin wave
excitations for the description of the ferromagnetic phase.
The order parameter, the magnetization, is the manifestation
of spontaneously broken rotation invariance and the spin waves
are collective modes whose energy goes to zero when the
wavelength goes to infinity.  These are  the  massless
Nambu-Goldstone bosons
\cite{nambu,nambujl,goldstone} associated to the spontaneously
broken symmetry.   Their dynamics essentially determine the
magnetization curve.  When the range of the forces between
spins is extended to all spins instead of, as is usually
considered, limited to near neighbors, the spin waves become
effectively massive and the magnetization curve encodes the
emergence of such a gap in the spectrum. This was the first
time we realized that the Nambu-Goldstone bosons, which signal
in general the spontaneous breakdown of a global symmetry,
cannot survive in presence of long range forces.

Our study of phase transition led us also to the analysis of
superconductivity.  We were extremely impressed by Nambu's
formulation of the BCS theory \cite{nambu2}. This paper and
the related   papers of Nambu and Jona-Lasinio \cite{nambujl}
on spontaneously broken chiral invariance   brought to light,
in full field theoretic terms,  the emergence of the massless
Nambu-Goldstone boson. They also pointed out the existence of
a massive scalar bound state  in the channel orthogonal to the
massless mode.   The significance of this massive scalar is
more transparent in the context of the Goldstone scalar field
model when the  symmetry breaking is driven by the  scalar
field potential itself \cite{goldstone}: it describes the
response to the order parameter. In the particular case of the
abovementioned ferromagnetic transition, this response is
the longitudinal susceptibility.   These beautiful papers were
certainly an element which later drove us into field theory.

In fall 1961, I was scheduled to return to Belgium. By that
time our collaboration   and our friendship   had become
deeply rooted. I received an offer of a professorship at
Cornell but I was missing Europe  very much. I decided not to
accept it and to return to Belgium. Robert and his wife
Martine had a similar attraction for the Old Continent; Robert
got a Guggenheim fellowship and they joined me in Belgium.
After a few months, the social life there and our personal
relations  decided Robert to resign from his professorship at
Cornell University and to settle permanently at Brussels
University.

We then  resumed in Belgium our analysis of broken symmetry.
We knew from our study of ferromagnetism that   long range
forces give mass to the spin waves   and we were aware,  from
Anderson's analysis of superconductivity \cite {anderson}, of
the fact  that the massless mode of neutral superconductors,
which is also a Nambu-Goldstone mode,   disappears in charged
superconductors in favor of the usual massive plasma
oscillations resulting from the long range coulomb
interactions in metals. Comforted by these facts, we   decided
to confront, in relativistic field theory, the long range
forces of Yang-Mills gauge fields with the   Nambu-Goldstone
bosons of a broken symmetry.

The latter arose from the breaking of a {\it global} symmetry
and Yang-Mills theory extends the symmetry to a {\it local}
one \cite {yang}.   Although the problem in this case is more
subtle   because of   gauge invariance,  the emergence of the
Nambu-Goldstone massless boson is very similar. We indeed
found that  there were well defined gauges in which the broken
symmetry induces such   modes.  But, as we expected, the long
range forces of the Yang Mills fields were conflicting with
those of the massless Nambu Goldstone fields. The conflict is
resolved by the generation of a mass reducing long range
forces to short range ones. In addition, gauge invariance
requires the Nambu-Goldstone mode to combine with the Yang
Mills excitations. In this way,   the gauge fields     acquire
a gauge invariant mass!

This work was finalized in 1964. We obtained the mass formula
for gauge fields, abelian and non abelian, where symmetry
breaking arises from non vanishing expectation values of
scalar fields \cite {be}. These play the role of ``order
parameters''. We also, guided by Nambu's work on
superconductivity, obtained through Ward identities a mass
formula for a dynamical symmetry breaking by fermion
condensate \cite{be}. Thus the  scalar fields  could be either
fundamental or constitute a phenomenological description of a
condensate.  Only future experimental and theoretical
development  can tell. Their massive excitations is   a
property of global symmetry breaking and are to be identified
with the  Goldstone (or with the Nambu) massive scalar bosons
describing the response to the ``order parameters''; they
are not altered by the introduction of local symmetry.  On the
contrary, the Nambu-Goldstone bosons which arise in group
space in directions  orthogonal to the  massive scalars, and
also  exist whether the scalar fields are fundamental or
composite, are ``eaten up''  by  the gauge fields which lay in
these directions. These acquire mass.  This fate of the
Nambu-Goldstone bosons is the  characteristic feature  of the
symmetry breaking mechanism in a local gauge symmetry.

I shall not dwell on subsequent developments but briefly
recall the most relevant steps. First there is the work of
Higgs who obtained   essentially the same results in a
somewhat different way \cite{higgs}.    He showed in simple
field theoretic terms   that the Nambu-Goldstone boson was
unobservable as such but  provided the required longitudinal
polarization for the gauge fields to get mass \cite{higgs2}.
This fact  is deeply related to the unitarity of the scheme
and is less explicit in our approach. On the other hand, our
formulation of the problem, using covariant gauges and Ward
identities, puts into evidence that   power counting for
Feynman graphs was consistent with  renormalizability.   This
is why we were led in 1966 to suggest that the theory of
vector mesons with mass generated  by the symmetry breaking
mechanism was renormalizable \cite{be2}.  But the full proof
of renormalizability was much more involved and in fact
required a detailed analysis of the consistency of the power
counting in covariant gauges and of the unitarity of the
theory. This  was worked out by Veltman and 't Hooft and the
proof was essentially completed in 1971
\cite{thooft}. It rendered the electroweak  theory
\cite{glashow,weinberg}, hitherto the most impressive
application of the mechanism  discovered by Higgs and
ourselves, a truly consistent and predictive scheme whose
experimental verification confirmed the validity of the
symmetry breaking mechanism.

This ends the story of the genesis of the unification scheme
which relates short and long range forces in gauge field
theory, indicating a path to the search for more general laws
of nature.  For us, it was only the beginning of our lasting
collaboration and of our lasting friendship.
\newpage

\noindent  {\bf SCIENTIFIC REVIEW }\footnote{Presented by R.
Brout}
\addtocounter{footnote}{-1}

The acquisition of mass by gauge vector mesons results from
the mutual coupling of  two fields each of which, in other
circumstances, had vanishing mass. These are, on one hand, the
zero mass excitations which result from the spontaneous
breakdown of symmetry (SBS) and, on the other, the zero mass
vector field which is necessitated when this same symmetry is
promoted from global to local, in which case it is called a
gauge symmetry. In relativistic theory these are called
respectively the Nambu-Goldstone (NG) and Yang-Mills (YM)
fields, although they have had a long prior history, the first
in statistical mechanics and solid state physics and the
second, of course, dating from Maxwell. I shall first explain
their masslessness in conceptual terms and then indicate how
their coupling gives rise to the vector mass.

We begin with the NG field which arises in consequence of SBS.
But firstly what is SBS? An amusing image has been given by
Abdus Salam.  Consider an ensemble of dinner companions seated
round the table set with plates between which is placed a
spoon. When the first guest chooses a spoon, to his right {\em
or} to his left, all others must follow suit. This is SBS.

The above example is a case of discrete SBS. A spoon on the
right (left) of a plate is represented by a spin which is on a
lattice site that is polarized up (down). Interaction favors
that neighboring spins be parallel. So the ground state is all
spins up (or down). An angel who fixes the polarization of
some spin then determines the polarization of all. Clearly it
costs a finite amount of energy to turn one spin against the
other so that this model presents an excitation spectrum which
has a gap.

Contrast this to SBS for the case of continuous symmetry. For
simplicity take the group $U(1)$. Then the spin has two
components $(S_x, S_y )$ such that $S_x^2+S_y^2 =1$. Once more
the ferromagnetic interaction favors neighbors to be parallel.
Imagine a state where all sit at an angle $\theta$ with
respect to the $x$-axis in group space. Clearly it costs no
energy to rotate them all through an angle
$\Delta\theta$, since they all remain parallel. The ground
state is thus degenerate with respect to $\theta$. Now divide
the system in two and rotate the two halves against each
other. Only the spins that ``rub'' against each other require
an energy to make such a configuration. Call this one node's
worth of energy. If one divides in thirds, it will cost two
nodes' worths of energy, and so on. Since translational
symmetry of the lattice requires that the excitations be
classed by wave number, $k$, we see that the energy grows with
$k$, and furthermore $\omega(k=0) =0$. So $k=0$ is the
terminal point of a spectrum which starts at zero frequency.
This is the NG excitation.

It is noteworthy that this process of the monotonic increase
of energy with the number of nodes stops once $k$ reaches the
inverse range of the force between spins, for then it cost no
more energy to make more nodes. Thus as the range tends to
infinity, the spectrum develops a gap, that is a mass, i.e.
$\lim_{k\to 0} \omega(k)$ is finite. This is a precursor of
what happens in general when long range forces are present.

I now explain why it is natural that gauge fields are also
massless. Take once more the simple case of spins. Global
symmetry is the invariance of the energy when all spins are
rotated ``en masse''. Local symmetry is realized when
different portions can be rotated differently (in group space)
at no cost of  energy. This is possible only if there is  a
``messenger'' which transmits from portion to portion the
information that such local rotations indeed do not cost any
energy and have no physical effect. In technical terms, this
messenger is called a connection or a gauge field. It
transforms under local rotations exactly in a way to
compensate for the
 energy that would otherwise follow from relative rotations of
neighboring spins. It is this beautiful idea which  governs
all the presently known interactions of nature.

From the above one understands that it is natural that the
gauge field has zero mass. Indeed under global transformations
a gauge field is not required to ensure invariance. So it
should not manifest itself. But a global transformation
corresponds to one whose wave vector $k_\mu$ vanishes: it
should cost  no energy $(k_0 =0)$ to make a gauge field
excitation which is everywhere the same $(\overline{k}=0)$. In
relativity the condition $k_\mu =0$ becomes the invariant
statement $k_\mu k^\mu =0$
$(\hbox{or}\  k_0^2 - \overline{k}^2 =0)$ which is the
statement of masslessness.

It is to be expected that a dramatic situation arises when
these two kinds of zero mass excitations are put
together in the context of a local symmetry. What happens is
that they combine into one massy vector field. The gauge
field, of itself, due to relativistic constraints has two
degrees of freedom. These are encoded in the  polarization
transverse to the direction of propagation. Massive vector
fields have a longitudinal polarization as well and this is
induced by the coupling to the NG field. To see how this
mechanism works it is convenient to express things in terms of
Feynman graphs.

In the case of no SBS, gauge fields propagate in vacuum by
taking into account the dielectric constant of the vacuum.
This is represented by loop insertion in the gauge field
propagator. For matter represented by a scalar field, a single
loop insertion is drawn as follows

\begin{center}
\bigskip
\leavevmode
\epsfbox {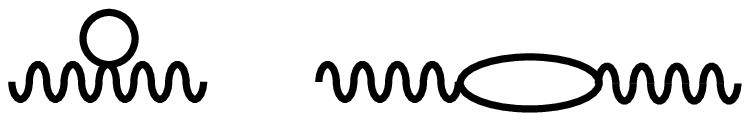}
\bigskip
\end{center}

These loops insertions in the YM propagator, represented by
wavy lines,  conserve the transverse character of the gauge
field and keep it massless. Their effect is to change the
value of the coupling constant of the gauge field to matter.

In the case of SBS, the finite expectation value of the scalar
field   causes additional graphs to arise which are found by
cutting the loop, generating the so-called tadpole graphs. One
must then addend to the above the graphs

\begin{center}
\bigskip
\leavevmode
\epsfbox {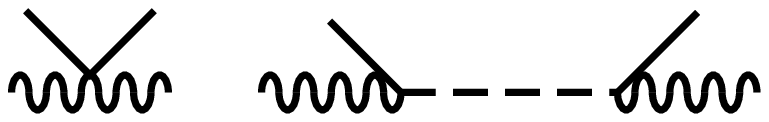}
\bigskip
\end{center}

In these graphs the wavy lines still represent the gauge field
and the solid line tadpoles are the expectation values of the
scalar fields which play the role of   order parameters.
The dashed line is the propagator of a NG excitation. These
arise in directions orthogonal in group space to the order
parameters.  The latter graphs   show how the NG field gets
absorbed into the gauge field, the net effect being to give to
the latter a mass proportional to  the order parameter and
to increase the number of degrees of freedom of the gauge
field from two to three. Although this ``order parameter'' is
here gauge dependent, there are Ward identities ensuring  the
gauge invariance of the mass arising from these graphs.  These
two elements are of utmost importance since the gauge
invariance ensures that the divergences of the graphs remain
under control, indicating that the theory could be
renormalizable, and the new longitudinal degree of freedom
renders the perturbation series unitary. It is the combination
of these two elements that   Veltman and 't Hooft used in
their     masterful works to prove  that the theory is indeed
renormalizable, thereby really setting the standard model on a
sound basis.

As just mentioned, the appearance of the massless NG bosons is
guaranteed by the Ward identities, and as such does not rely on
perturbation theory. They   therefore also appear if SBS is
realized dynamically through a fermion condensate, as in the
BCS theory of superconductivity or in the Nambu Jona-Lasinio
theory of broken global chiral symmetry. In presence of a
local symmetry, they would then still generate a mass for the
gauge vector mesons. In that case, the scalar fields would
be  phenomenological rather than fundamental objects but the
mechanism would remain essentially the same.  Whether
fundamental or not, the scalar fields describing the order
parameters, have massive quanta. These massive scalars  are
not  a specific feature of the mechanism: they arise already
in global SBS, and even in discrete ones such as our original
discrete spin system. The free energy of such a system
presents as a function of the magnetization, below the Curie
point, the double dip shape typical of the Landau-Ginsburg
potential $V$ represented in the figure below. This potential
is the same as the one driving global SBS in the Goldstone
scalar field model. The distance of the dip to the origin and
the curvature at this point are respectively  the expectation
value and the mass squared of the Goldstone massive scalar
boson. The latter, or more precisely its inverse mass squared,
measures  the longitudinal susceptibility. This is the
response of a field parallel to the order parameter and
appears in any second order phase transition.

\begin{center}
\bigskip
\leavevmode
\epsfbox {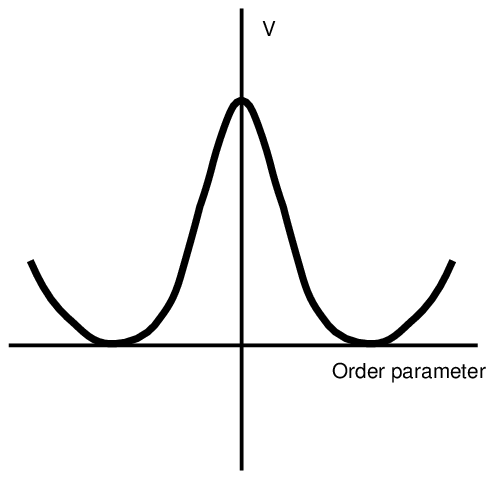}
\bigskip
\end{center}

I conclude by restating the main results with  some emphasis
on  their phenomenological implications.

Massive gauge vector mesons are an inevitable consequence of
SBS, independently of the dynamical mechanism which causes the
breaking, scalar fields\footnote{Within the past few years,
an interesting development has occurred principally due to
the mathematician A. Connes who has applied techniques of non
commutative geometry to construct the standard model. In this,
the key point is that the scalar field plays the role of a
gauge connection in the ``motion'' of a fermion (whose mass is
generated by spontaneously broken chiral symmetry) during its
zitterbewegung.}, bound state condensate... Observationally
in the electroweak sector they occur as narrow resonances
because their coupling to the continuum is small. This latter
is an observational fact: the electric charge is small and thus
governs the scale of all gauge field couplings.

Massive scalar occur in channels orthogonal to the NG
channels. They appear in any global SBS and are thus not a
specific feature of mass generation for gauge fields;   their
physics is accordingly much more sensitive to dynamical
assumptions. They too would appear as resonances whether or
not they be manifestations of an elementary scalar field or as
composites due to a more elaborate mechanism. Whether or not
these resonances are swamped into the continuum depends on the
parameters of the theory and we do not control these in the
same way as we control coupling to the gauge vectors. Thus
observation of both mass and width in these channels will
deliver to us precious indications of the mechanism at work.

In all cases SBS  in   gauge theories is characterized by
NG  bosons   which are ``eaten up'' by the YM fields, giving
them longitudinal polarization and  mass. It is this
phenomenon which allows for consistent renormalizable theories
of massive gauge vector mesons.

\end{document}